\documentclass{article}
\usepackage{spconf,amsmath,graphicx,hyperref}
\usepackage{amssymb}  %
\usepackage{booktabs}
\usepackage{multirow}
\usepackage{makecell}

\usepackage{array}

\title{CLAIP-Emo: Parameter-Efficient Adaptation of Language-supervised models for In-the-Wild Audiovisual Emotion Recognition}

\name{Yin Chen, Jia Li, Jinpeng Hu,  Zhenzhen Hu, Richang Hong} %
\address{Hefei University of Technology, Hefei, China}
\begin{document}
\maketitle

\begin{abstract}
	\ninept

	Audiovisual emotion recognition (AVER) in the wild is still hindered by pose variation, occlusion, and background noise. Prevailing methods primarily rely on large-scale domain-specific pre-training, which is costly and often mismatched to real-world affective data. To address this, we present CLAIP-Emo, a modular framework that reframes in-the-wild AVER as a parameter-efficient adaptation of language-supervised foundation models (CLIP/CLAP). Specifically, it (i) preserves language-supervised priors by freezing CLIP/CLAP backbones and performing emotion-oriented adaptation via LoRA (updating  \ensuremath{\le}4.0\% of the total parameters), (ii) allocates temporal modeling asymmetrically, employing a lightweight Transformer for visual dynamics while applying mean pooling for audio prosody, and (iii) applies a simple fusion head for prediction. On DFEW and MAFW, CLAIP-Emo (ViT-L/14) achieves 80.14\% and 61.18\% weighted average recall with only 8M training parameters, setting a new state of the art. Our findings suggest that parameter-efficient adaptation of language-supervised foundation models provides a scalable alternative to domain-specific pre-training for real-world AVER.  The code and models will be available at \href{https://github.com/MSA-LMC/CLAIP-Emo}{https://github.com/MSA-LMC/CLAIP-Emo}.
\end{abstract}

\section{Introduction}
\label{sec:intro}
\ninept

Recognizing human emotions in the wild from audiovisual cues (AVER) is a cornerstone of affective computing~\cite{pantic2002automatic}, yet it remains a formidable challenge due to uncontrolled environmental factors such as unpredictable lighting, pose variations, occlusions, and diverse acoustic noise~\cite{schuller2011avec, dhall2015video}.  The dominant paradigm to tackle this involves a two-stage process:  large-scale self-supervised pre-training (e.g., MAE~\cite{he2022mae}) on domain-specific corpora like VoxCeleb2~\cite{chung2018voxceleb2}, which contains human faces and voices, followed by full fine-tuning on the target emotion dataset~\cite{sun2024hicmae,cheng2025vaemo,wu2025avf}. While effective, this approach  suffers from two fundamental limitations: (i) \textbf{high computational costs}, as both pre-training and full fine-tuning require substantial resources, slowing research iteration and limiting deployment; and (ii) \textbf{potential semantic gap}, as self-supervised objectives tend to capture holistic representations of the input modality, rather than focusing on the subtle, componential cues that constitute emotional expression.

\begin{figure}[!h]
	\centering
	\ninept
	
	\includegraphics[width=\linewidth]{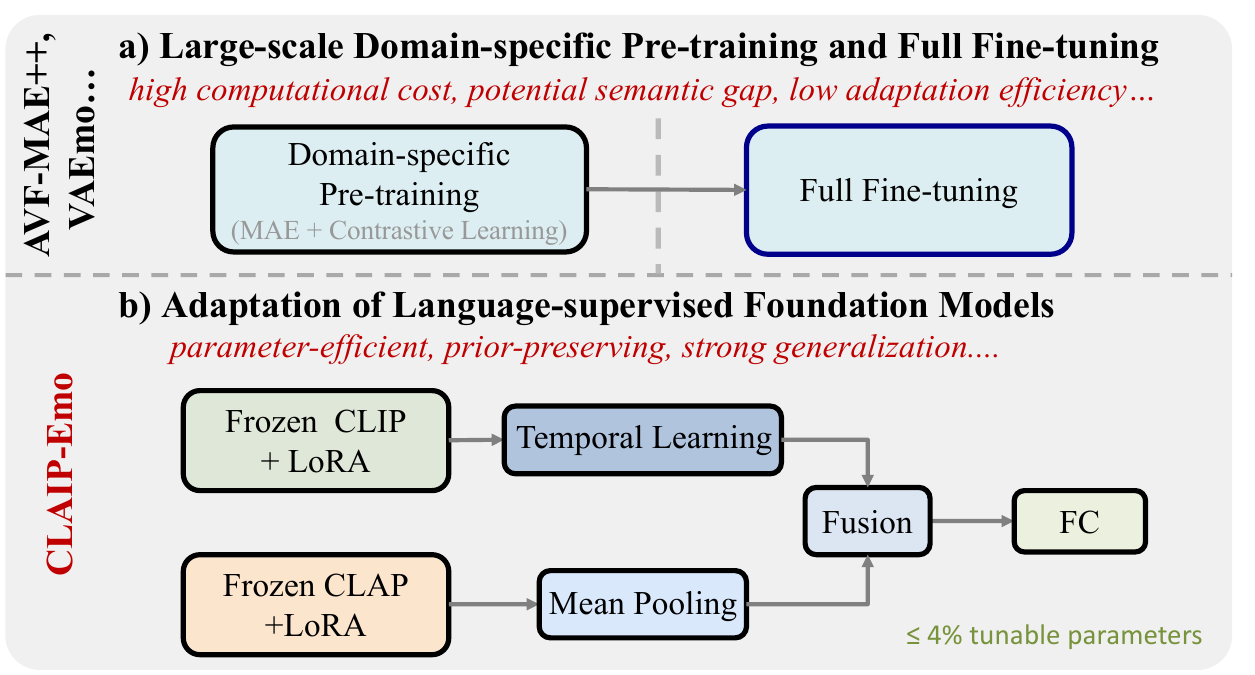} %
	\caption{Domain-specific pre-training and full fine-tuning vs. our parameter-efficient foundation model adaptation (CLAIP-Emo).}
	\label{fig:pipeline}
	\vspace{-10pt}
\end{figure}

\begin{figure*}[htbp]
	\ninept
	
	\centerline{
		\includegraphics[width=\linewidth]{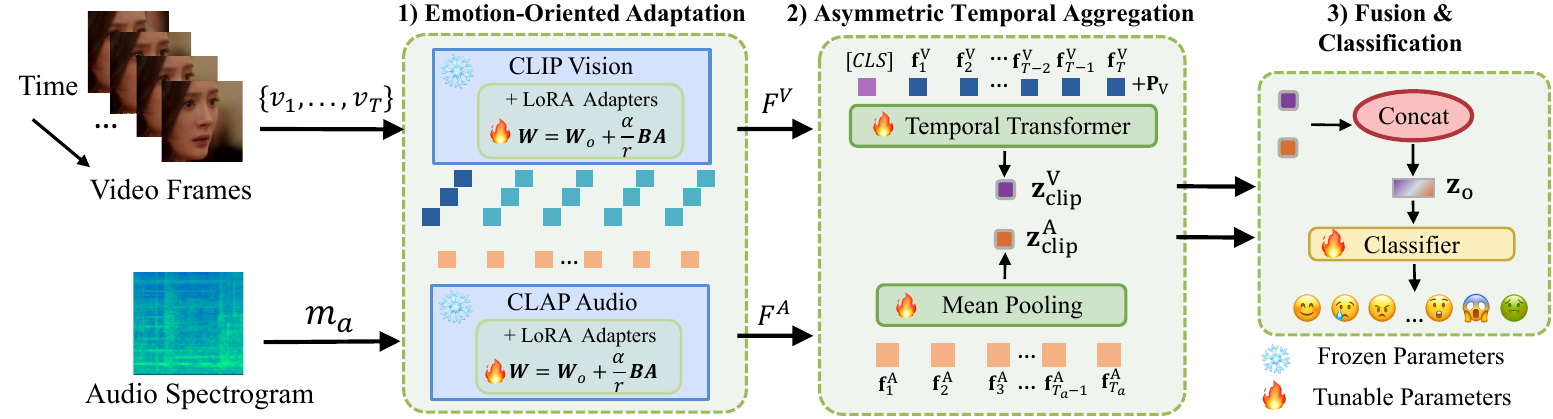}}
	\caption{\textbf{Overview of the CLAIP-Emo framework.} Our model adapts frozen CLIP and CLAP backbones using lightweight LoRA adapters. An asymmetric temporal module processes visual and audio dynamics differently, before a simple fusion head predicts the final emotion.}
	\label{fig:framework}
\end{figure*}

Language-supervised Foundation Models (LFMs), such as the vision-language model CLIP \cite{radford2021learning} and the audio-language model CLAP \cite{elizalde2023clap}, offer a compelling alternative.  Pre-trained on web-scale data to align raw signals with natural language descriptions, these models acquire semantically rich and robust representations of the world. This linguistic grounding provides a more direct pathway to \textit{interpreting the specific, componential cues that constitute high-level concepts like affective states},  thus offering a promising foundation for AVER without requiring costly domain-specific pre-training. 

Despite their potential, adapting these powerful LFMs for AVER poses two fundamental challenges. First, conventional full fine-tuning, while common in unimodal adaptation~\cite{zhao2023dferclip,li2024domain}, risks catastrophic forgetting~\cite{he2021analyzing} by aggressively updating entire parameters, thereby \textit{corrupting the semantic priors  crucial for affective understanding and leading to overfitting on limited emotion data.} Second, and more subtly, the inherent nature of CLIP and CLAP presents an \textit{asymmetry dilemma}: CLIP's visual encoder, pre-trained on static images, excels at spatial representation but is blind to the temporal dynamics of emotional expressions. In contrast, CLAP's audio encoder, pre-trained on entire audio clips, innately captures holistic, clip-level vocal semantics. A naive, symmetric adaptation would either fail to build necessary temporal awareness for visual emotion cues or disrupt the powerful, pre-existing global prior for auditory sentiment.

To resolve these challenges, we introduce CLAIP-Emo (Fig.~\ref{fig:pipeline}, \ref{fig:framework}), a novel framework based on the core principle that \textit{adaptation must respect the inherent properties of the foundation models.} This principle manifests in a two-fold strategy. First, to preserve the rich, pre-trained semantic priors and prevent catastrophic forgetting, we freeze the CLIP and CLAP backbones and employ lightweight LoRA~\cite{hu2022lora} adapters. This allows for task-specific specialization by updating a mere fraction (\ensuremath{\le}4.0\%) of the total parameters. Second, we introduce a deliberately asymmetric temporal aggregation module to align with each model's distinct architectural priors.  For the vision stream, which lacks temporal context, we introduce a lightweight Transformer to model the dynamic evolution of expressions from frame-level features explicitly. In contrast, for the audio stream, we employ simple mean-pooling to leverage and preserve the holistic, clip-level representation intrinsic to the CLAP encoder. These tailored representations are then fused via a simple ``concat+linear" head for final prediction. This principled design leads to a simple yet powerful architecture, eliminating the need for complex cross-modal alignment mechanisms.

Our main contributions are threefold: 1) we propose a parameter-efficient, prior-preserving adaptation framework that enables LFMs for AVER while bypassing the need for costly domain-specific pre-training; 2) we introduce CLAIP-Emo, a novel architecture that exploits the asymmetric strengths of CLIP and CLAP through LoRA-based tuning and a temporal aggregation module aligned with their visual-spatial and audio-holistic priors; 3) with only 8M tunable parameters, CLAIP-Emo sets new state-of-the-art performance on DFEW~\cite{jiang2020dfew} (80.14\% Weighted Average Recall, WAR) and MAFW~\cite{liu2022mafw} (61.18\% WAR), establishing a simple, reproducible, and powerful baseline.

\section{Methodology}
\label{sec:method}
We propose \textbf{CLAIP-Emo},  a parameter-efficient and prior-preserving framework for adapting foundation models to audiovisual emotion recognition (AVER). Our end-to-end architecture (Fig.~\ref{fig:framework}) first adapts frozen CLIP/CLAP encoders via LoRA. It then applies asymmetric temporal aggregation tailored to each modality's dynamics before fusing the representations with a simple fusion head for final classification. This design ensures both high performance and efficiency.

\subsection{Problem Formulation} \label{ssec:problem_formulation}

Given a video clip $C=(\mathcal{V},A)$ consisting of $T$ visual frames  $\mathcal{V} = \{v_1,\dots,v_T\}$ and its corresponding audio waveform $A$, the AVER task is defined as predicting an emotion label $\hat{y}$ from a predefined category set $\mathcal{C}=\{c_1,\dots,c_K\}$ by maximizing the conditional probability $P(y|\mathcal{V},A)$, where  $K$ is the number of emotion classes and $y\in\mathcal{C}$ denotes the ground-truth label. Our framework $\Phi\colon(\mathcal{V},A)\mapsto \hat{y}$ is optimized end-to-end using cross-entropy loss.

\subsection{Emotion-Oriented Parameter-Efficient Adaptation}
\label{sec:adaptation}
A central challenge in adapting foundation models is to specialize them for a downstream task without suffering from catastrophic forgetting or corrupting their rich, language-aligned priors. To address this, we adopt a Parameter-Efficient Fine-Tuning (PEFT) strategy using Low-Rank Adapters (LoRA)~\cite{hu2022lora}. Instead of updating the entire model, we freeze the original weights of the CLIP vision (ViT) and CLAP audio (HTSAT) encoders, and inject lightweight, trainable LoRA modules into each attention and MLP layer. Formally, LoRA modifies a frozen weight matrix $\mathbf{W}_0
	\in\mathbb{R}^{d_{\text{out}}\times d_{\text{in}}}$ by adding a low-rank update:
\begin{equation}
	\mathbf{W} = \mathbf{W}_0 + \frac{\alpha}{r}\,\mathbf{B}\mathbf{A},\quad
	\mathbf{B}\in\mathbb{R}^{d_{\mathrm{out}}\times r},\;
	\mathbf{A}\in\mathbb{R}^{r\times d_{\mathrm{in}}},
	\label{eq:lora}
\end{equation}
where only the matrices $A$ and $B$ are trainable. We set $r=8$ and $\alpha=32$, yielding a tunable fraction of $4.0\%$ for ViT-B/16 and $2.5\%$ for ViT-L/14 with CLAP-HTSAT, enabling efficient adaptation while preserving foundation-model priors.

\textbf{Visual branch.}
Let $\mathcal{E}_\mathrm{V}(\cdot;\theta_\mathrm{V},\phi_\mathrm{V})$ denote the adapted CLIP encoder with frozen backbone $\theta_\mathrm{V}$ and trainable LoRA parameters $\phi_\mathrm{V}$. Each sampled visual frame $v_t$ is independently encoded, and we use the \texttt{[CLS]} token as the frame representation, i.e., $\mathbf{f}_t^{\mathrm{V}} = \mathcal{E}_\mathrm{V}(v_t;\theta_V, \phi_V)_{[0]}$, yielding $\mathbf{F}^{\mathrm{V}} = [\,\mathbf{f}_1^{\mathrm{V}},\ldots,\mathbf{f}_T^{\mathrm{V}}\,]^{\mathsf{T}} \in \mathbb{R}^{T\times d_{\mathrm{V}}}$.

\textbf{Audio branch.}
Similarly, we adapt the CLAP audio encoder $\mathcal{E}_\mathrm{A}(\cdot;\theta_\mathrm{A},\phi_\mathrm{A})$. The waveform $A$ is converted to a Mel spectrogram $m_a\in\mathbb{R}^{T_a \times F_a}$, where $T_a$ is the number of time frames and $F_a$ is the number of Mel frequency bins.  It is then encoded as a  sequence, giving $\mathbf{F}^\mathrm{A} = \mathcal{E}_\mathrm{A}(m_a;\theta_\mathrm{A},\phi_\mathrm{A})\in\mathbb{R}^{T_a\times d_\mathrm{A}}$.

\subsection{Asymmetric Temporal Aggregation}
\label{sec:aggregation}
We adopt an asymmetric temporal aggregation strategy aligned with the pretraining objectives and feature granularity of each encoder. The visual branch employs an image-based CLIP encoder, which produces frame-level features without modeling cross-frame context, whereas the audio branch leverages CLAP, trained with a clip-level contrastive objective to capture global acoustic information. Accordingly, we introduce branch-specific asymmetric temporal aggregation, which compresses $\mathbf{F}^\mathrm{V}$ and $\mathbf{F}^\mathrm{A}$ into compact clip-level representations while maintaining a balance between representational capacity and computational efficiency.

\textbf{Modeling Visual Dynamics.}
Visual emotional expressions are inherently dynamic, defined by the temporal evolution of facial cues. To capture these inter-frame dependencies, we model the visual sequence $\mathbf{F}^\mathrm{V} $ with a lightweight temporal Transformer. We prepend a learnable \texttt{[CLS]} token and add positional embeddings $\mathbf{P}_\mathrm{V} \in \mathbb{R}^{(T+1) \times d_\mathrm{V}}$ to form the input:
\begin{equation}
	\label{eq:transformer_input}
	\mathbf{Z}_0^\mathrm{V} = [\,\texttt{[CLS]}; \mathbf{f}_1^\mathrm{V}; \dots; \mathbf{f}_T^\mathrm{V}\,] + \mathbf{P}_\mathrm{V}.
\end{equation}
After processing by a single Transformer layer, the output embedding of the \texttt{[CLS]} token, which aggregates sequence-wide information, is taken as the final clip-level visual feature:
\begin{equation}
	\label{eq:transformer_output}
	\mathbf{z}_\mathrm{clip}^\mathrm{V} = \text{Transformer}(\mathbf{Z}_0^\mathrm{V})_{[0]}.
\end{equation}

\textbf{Preserving Audio Priors.}
In contrast, the CLAP audio encoder is optimized for clip-level summarization under a contrastive pretraining objective. We thus avoid adding heavy temporal heads on top of the audio sequence $\mathbf{F}^\mathrm{A}$ and instead apply simple mean pooling:
\begin{equation}
	\label{eq:audio_pooling}
	\mathbf{z}_\mathrm{clip}^\mathrm{A} = \frac{1}{T_a} \sum_{t=1}^{T_a} \mathbf{f}_t^\mathrm{A}.
\end{equation}
This operation is not only efficient but also preserves CLAP’s language-supervised priors, yielding a robust and holistic representation of the clip’s acoustic content.

Finally, the asymmetric design yields two compact representations, $\mathbf{z}_\mathrm{clip}^\mathrm{V}$ and $\mathbf{z}_\mathrm{clip}^\mathrm{A}$, which are subsequently fused for multimodal emotion understanding.

\subsection{Lightweight  Fusion and Classification}

A key principle of our design is that effective modality-specific processing obviates the need for complex fusion.
Building upon the tailored representations ($\mathbf{z}_\mathrm{clip}^\mathrm{V}, \mathbf{z}_\mathrm{clip}^\mathrm{A}$) from the preceding stage, we therefore employ a minimalist fusion head. The features are simply concatenated and then projected by a linear classifier for final prediction:
\begin{align}
	\mathbf{z}_{\mathrm{o}} &= \text{Concat}(\mathbf{z}_\mathrm{clip}^\mathrm{V}, \mathbf{z}_\mathrm{clip}^\mathrm{A}), \\
	\hat{\mathbf{p}} &= \text{Softmax}(\mathbf{W}_c \mathbf{z}_{\mathrm{o}} + b_c),
\end{align}
where $\mathbf{W}_c \in \mathbb{R}^{K \times (d_V + d_A)}$ and $\mathbf{b}_c \in \mathbb{R}^{K}$ are the learnable parameters for the $K$ emotion classes.  Here, $\hat{\mathbf{p}}$ denotes the predicted class probabilities, and the final predicted label is obtained as $\hat{y} = \arg\max \hat{\mathbf{p}}$.

\section{Experiments}
We evaluate CLAIP-Emo on the DFEW~\cite{jiang2020dfew} and MAFW~\cite{liu2022mafw} datasets. Audio preprocessing follows the CLAP pipeline~\cite{elizalde2023clap}, while video preprocessing adopts S4D~\cite{chen2024s4d}. We present two variants: CLAIP-Emo-B (ViT-B/16) and CLAIP-Emo-L (ViT-L/14), containing $4.0\%$ and $2.5\%$ trainable parameters, respectively. Both employ the CLAP audio encoder, with the B-variant used as the default for ablations. LoRA adapters are configured with $r{=}8$, $\alpha{=}32$, and dropout $0.1$. Training is performed for 100 epochs on two NVIDIA RTX A40 GPUs using Adam with a cosine learning-rate schedule (5-epoch warmup), batch size 16 per GPU, and an initial learning rate of $1\times 10^{-5}$. We report mean Unweighted Average Recall (UAR) and Weighted Average Recall (WAR) across the official five-fold splits.

\begin{table}[t]
	\centering
	\ninept
	
	\setlength{\tabcolsep}{3pt}  %
	
	\caption{Ablations on DFEW (fold 1) / MAFW (fold 1). Metrics: UAR / WAR (\%). ``Tunable'' reports trainable parameters in millions and ratio. Trans.: Transformer; Mean: mean pooling; A: Audio; V: Video. Best in \textbf{bold}. fd1: fold 1.}
	\begin{tabular}{l|c|c|r}
		\hline
		
		Variant                                    & \makecell{DFEW (fd1)                                                          \\UAR/WAR}                             & \makecell{MAFW (fd1) \\UAR/WAR}                                         & \makecell{Tunable \\M / \%}  \\
		
		\hline

		\multicolumn{4}{l}{\textit{LoRA Adaptation}}                                                                               \\
		\quad Frozen  ($r{=}0$)                    & 59.99 / 74.30                   & 32.78 / 46.68                   & 3.2 / 2.7 \\
		\quad LoRA ($r{=}2$)                       & 64.70 / 78.30                   & 36.63 / 50.65                   & 3.6 / 3.0 \\
		\quad LoRA ($r{=}4$)                       & 65.07 / 78.50                   & 37.24 / 50.65                   & 4.1 / 3.4 \\
		\quad \textbf{LoRA ($\boldsymbol{r{=}8}$)} & \textbf{65.96} / \textbf{79.84} & \textbf{37.34} / \textbf{51.14} & 5.0 / 4.0 \\
		\quad LoRA ($r{=}16$)                      & 65.73 / 78.90                   & 37.27 / 50.98                   & 6.7 / 5.4 \\
		\quad Full Fine-tune                       & 60.40 / 74.08                   & 32.13 / 42.43                   & 119 / 100 \\
		
		\hline
		
		\multicolumn{4}{l}{\textit{Temporal Aggregation Strategy}}                                                                 \\
		\quad V: Mean / A: Mean                    & 64.46 / 77.11                   & 36.38 / 49.56                   & 1.8 / 1.5 \\
		\quad \textbf{V: Trans. / A: Mean}         & \textbf{65.96} / \textbf{79.84} & \textbf{37.34} / \textbf{51.14} & 5.0 / 4.0 \\
		\quad V: Trans. / A: Trans.                & 64.61 / 78.82                   & 36.09 / 51.09                   & 8.1 / 6.5 \\
		
		\hline
		
		\multicolumn{4}{l}{\textit{Fusion Head}}                                                                                   \\
		\quad Additive Fusion                      & 64.08 / 78.65                   & 35.17 / 50.11                   & 5.0 / 4.1 \\
		\quad Gated Fusion                         & 65.79 / 79.24                   & 38.48 / 50.76                   & 5.5 / 4.5 \\
		\quad \textbf{Concat + Linear}             & \textbf{65.96} / \textbf{79.84} & \textbf{37.34} / \textbf{51.14} & 5.0 / 4.0 \\
		
		\hline
		
	\end{tabular}
	\label{tab:ablation_dual}
\end{table}
\subsection{Ablation Studies}
\label{sec:experiments}

\textbf{Effectiveness of Model Components.}
We conducted ablation studies to assess the contributions of various components within CLAIP-Emo, with results summarized in Table~\ref{tab:ablation_dual}. 
Increasing the rank from $r{=}0$ (frozen backbones) to $r{=}8$ lifts WAR by +5.54 points on DFEW (74.30\%→79.84\%) and +4.46 points on MAFW (46.68\%→51.14\%) while tuning only 5.0\,M (4.0\%) parameters. 
Further increasing $r$ to $16$ slightly degrades performance (e.g., 79.84\%→78.90\% on DFEW), and full fine-tuning of all parameters reduces DFEW WAR to 74.08\% and MAFW WAR to 42.43\%, which confirms the importance of preserving foundation-model priors.  
For temporal modeling, a single Transformer layer outperforms frame mean pooling by $+2.73$\% WAR on DFEW and $+1.58$\% WAR on MAFW. Upgrading the audio stream from mean pooling to a Transformer increases trainable parameters by about $62\%$ with negligible gains, which supports our asymmetric design. For multimodal fusion, a minimalist ``Concat+Linear" head yields the best trade-off. \textit{Additive fusion reduces accuracy and gated fusion brings no benefit on MAFW despite higher complexity.}
In summary, tuning 4\% parameters of the backbone, modeling only visual dynamics, and employing a linear fusion head collectively provide CLAIP-Emo with an optimal balance between performance and efficiency.

\begin{table}[t]
	
	\centering
	\ninept
	\setlength{\tabcolsep}{4pt}
	\caption{Ablation on modality and backbone. Metrics: UAR / WAR (\%). Visual backbone: CLIP ViT-B/16; audio backbone: CLAP HTSAT. fd1: fold 1.}
	\begin{tabular}{l|c|c|c}
		\hline
		
		Modality & Backbone         & \makecell{DFEW (fd1)                 \\UAR/WAR}          & \makecell{MAFW (fd1) \\UAR/WAR}          \\
		\hline
		
		A        & HTSAT            & 38.15 / 47.52        & 20.92 / 30.88 \\
		V        & ViT-B/16         & 62.60 / 76.60        & 34.61 / 49.24 \\
		A+V      & ViT-B/16 + HTSAT & 65.96 / 79.84        & 37.34 / 51.14 \\
		\hline
		
	\end{tabular}
	\label{tab:modality_backbone}
\end{table}

\textbf{Modality Contribution.}
Table~\ref{tab:modality_backbone} quantifies the contribution of each modality. While the audio-only model provides a reasonable baseline (e.g., 38.15\%/47.52\% UAR/WAR on DFEW), the visual model serves as the primary contributor, achieving a significantly higher UAR/WAR of 62.60\%/76.60\%. \textit{This trend is consistent on MAFW, underscoring the primacy of facial cues for in-the-wild emotion recognition.} Ultimately, fusing both modalities yields the best results on both datasets (65.96\%/79.84\% on DFEW and 37.34\%/51.14\% on MAFW), demonstrating that audio provides crucial complementary information and validating the synergistic power of the CLIP and CLAP backbones.

\begin{table}[t]
	\centering
	\ninept	
	\caption{Ablation on visual pre-training priors. The audio branch remains unchanged. Results are reported on DFEW and MAFW. fd1: fold 1.}
	\begin{tabular}{l|c|c}
		\hline
		
		Visual Backbone Pre-training & \makecell{DFEW (fd1)                 \\UAR/WAR}  & \makecell{MAFW (fd1)  \\UAR/WAR}                       \\
		\hline
		Random Initialization        & 39.86 / 48.91        & 21.37 / 30.99 \\
		ImageNet-21k (Supervised)    & 59.41 / 72.92        & 31.75 / 45.97 \\
		CLIP-ViT (Ours)              & 65.96 / 79.84        & 37.34 / 51.14 \\
		
		\hline
		
	\end{tabular}
	\label{tab:ablation_priors}
\end{table}

\textbf{Visual Prior Ablation.} We investigate the impact of visual pre-training in Table~\ref{tab:ablation_priors} by replacing the CLIP-ViT backbone while keeping the CLAP audio branch fixed. As expected, adapting a randomly initialized ViT yields poor results (e.g., 48.91\% WAR on DFEW), confirming that strong pre-trained priors are indispensable. While a standard ImageNet-pretrained ViT provides a competitive baseline, our CLIP-based model significantly outperforms it,  yielding +6.9\% and +5.17\% WAR gains on DFEW and MAFW, respectively. These results underscore the superiority of CLIP's language-aligned semantic priors over ImageNet's object-centric ones for emotion recognition.
\vspace{-6pt}  %

\begin{table}[t]
	\centering
	\ninept
	
	\setlength{\tabcolsep}{3pt}  %
	
	\caption{Comparison with state-of-the-art methods on in-the-wild AVER. Metrics: UAR / WAR (\%). Mod.: modality; TP.: tunable parameters (M). Best in \textbf{bold}, second best \underline{underlined}.}
	\begin{tabular}{l|c|c|c|c}
		\hline
		
		Method                            & Mod. & TP. & \makecell{DFEW                                                          \\UAR/WAR}  & \makecell{MAFW \\UAR/WAR}                       \\
		\hline
		
		HuBERT~\cite{hsu2021hubert}       & A    & 95   & 36.95 / 43.24                   & 25.00 / 32.60                         \\
		WavLM-Plus~\cite{chen2022wavlm}   & A    & 95   & 37.78 / 44.64                   & 26.33 / 34.07                         \\
		MAE-DFER \cite{sun2024hicmae}     & V    & 85   & 63.41 / 74.43                   & 41.62 / 54.31                         \\
		DFER-CLIP \cite{zhao2023dferclip} & V    & -    & 59.61 / 71.25                   & 39.89 / 52.59                         \\
		DK-CLIP \cite{li2024domain}       & V    & -    & 64.95 / 75.41                   & 43.01 / 56.56                         \\
		S2D~\cite{chen2024s2d}            & V    & 9    & 61.82 / 76.03                   & 41.86 / 57.37                         \\
		S4D~\cite{chen2024s4d}            & V    & 101  & 66.80 / 76.68                   & 43.72 / 58.44                         \\
		HiCMAE-B \cite{sun2024hicmae}     & AV   & 81   & 63.76 / 75.01                   & 42.65 / 56.17                         \\
		VAEmo \cite{cheng2025vaemo}       & AV   & 39   & 64.02 / 75.78                   & 45.67 / 58.91                         \\
		AVF-MAE++ (B) \cite{wu2025avf}    & AV   & 169  & 63.74 / 76.24                   & 43.10 / 57.50                         \\
		AVF-MAE++ (L) \cite{wu2025avf}    & AV   & 303  & 65.14 / 75.42                   & 45.36 / 59.13                         \\
		AVF-MAE++ (H) \cite{wu2025avf}    & AV   & 521  & \underline{66.88} / 77.45       & 46.05 / 60.24                         \\
		\hline
		
		CLAIP-Emo (ViT-B/16)              & AV   & 5    & 66.21 / \underline{78.42}       & \underline{45.58} / \underline{60.44} \\
		CLAIP-Emo (ViT-L/14)              & AV   & 8    & \textbf{69.52} / \textbf{80.14} & \textbf{46.65} / \textbf{61.18}       \\
		\hline
		
	\end{tabular}
	\label{tab:sota_inwild}
\end{table}

\subsection{Comparison with State-of-the-Art Methods}
\begin{figure}[!ht]
	\centering
	\ninept
	
	\includegraphics[width=0.95\linewidth]{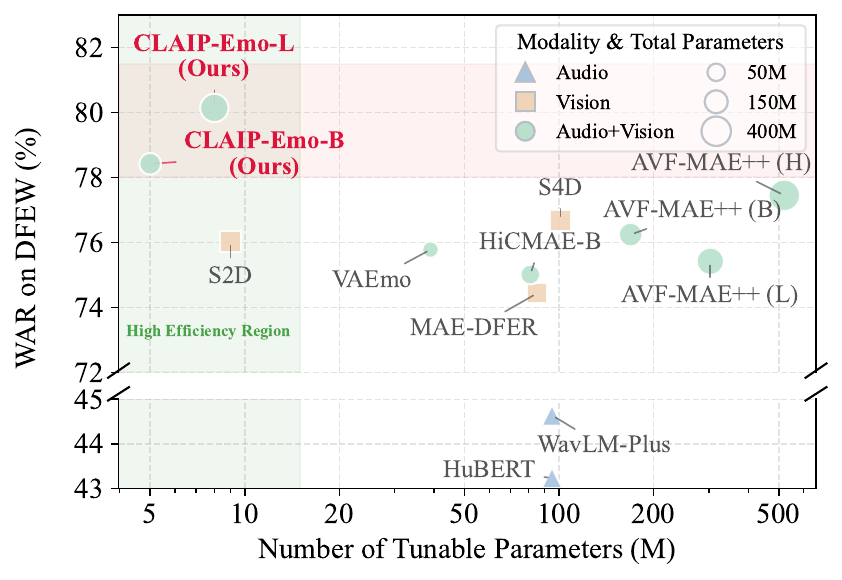} %
	
	\caption{CLAIP-Emo achieves state-of-the-art performance with significantly fewer tunable parameters on DFEW.}
	
	\label{fig:sota}
\end{figure}

We compare CLAIP-Emo with recent state-of-the-art methods in Table~\ref{tab:sota_inwild}. Our smaller model, CLAIP-Emo-B (ViT-B/16), \textit{with only 5M tunable parameters, already surpasses all previous work in WAR on both benchmarks.} Notably, it outperforms the much larger AVF-MAE++(H) (521M) on DFEW and MAFW by +0.97\% and +0.20\% WAR, respectively, while having a comparable UAR.
\textit{By scaling the visual backbone to ViT-L/14, our CLAIP-Emo-L model (8M tunable parameters) establishes a new state of the art.} It surpasses the previous best method on DFEW by +2.64\%/+2.69\% UAR/WAR, and on MAFW by +0.60\%/+0.94\% UAR/WAR. This remarkable performance is achieved while tuning less than 2\% of the parameters of AVF-MAE++(H). This superior trade-off between performance and  parameter efficiency is visualized in Fig.~\ref{fig:sota}, where CLAIP-Emo occupies the top-left corner, signifying the best performance with minimal tunable parameters. Furthermore, compared to the strongest vision-only model, S4D (101M), CLAIP-Emo-L shows significant gains across all metrics (e.g., +3.46\% WAR on DFEW). These results provide strong evidence that parameter-efficient adaptation of language-supervised models can outperform heavily pre-trained, large-scale pipelines on challenging in-the-wild AVER tasks.

\begin{figure}[h]
	\centering
	\includegraphics[width=\linewidth]{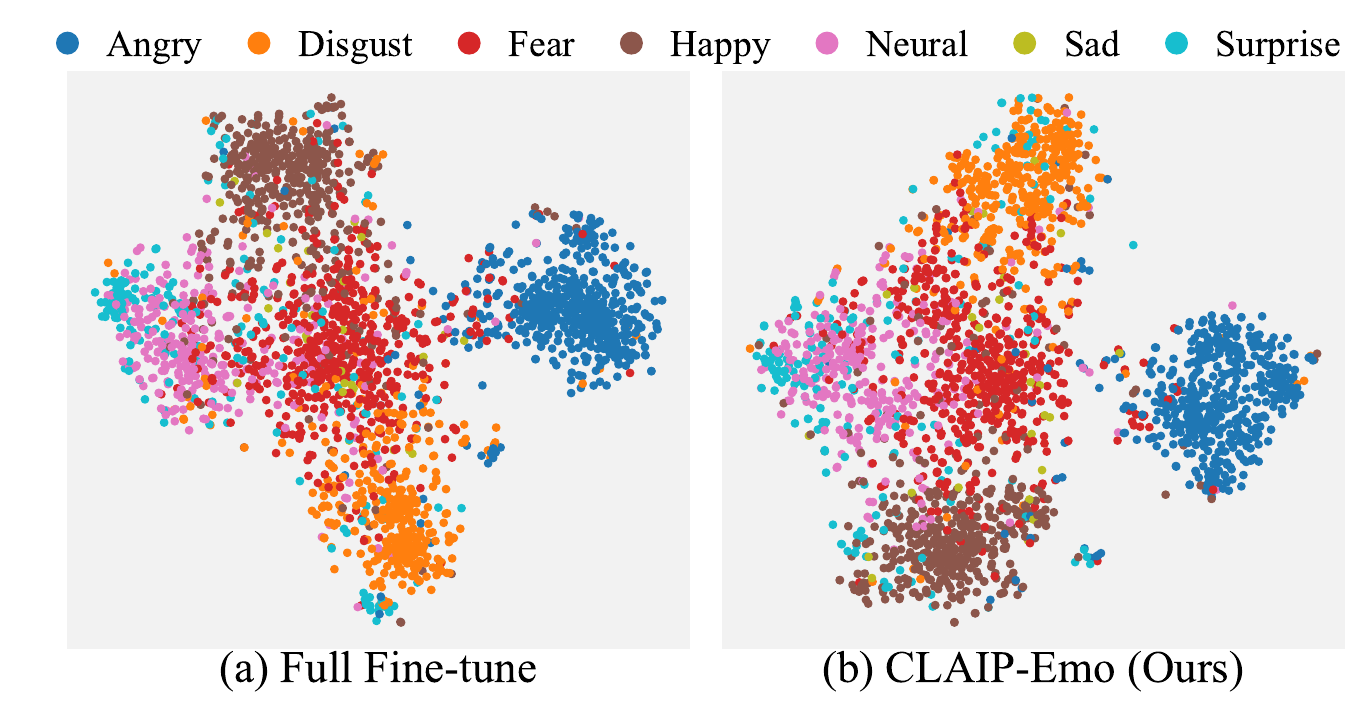} %
	\caption{t-SNE~\cite{maaten2008visualizing} visualization of features extracted by (a) full fine-tuning and (b) CLAIP-Emo on DFEW fold 1.}
	\label{fig:tsne}
\end{figure}

\subsection{Feature Visualization}
Figure~\ref{fig:tsne} visualizes the learned feature spaces. The full fine-tuning baseline (a) suffers from severe cluster overlap, particularly for classes like \textit{Fear} and \textit{Surprise}, indicating feature confusion. In contrast, CLAIP-Emo (b) demonstrates a markedly improved feature space, yielding more compact and better-defined clusters overall. This enhanced feature discriminability, characterized by reduced intra-class variance and increased inter-class margins, helps explain our quantitative gains and validates the effectiveness of our prior-preserving adaptation strategy.

\vspace{-5pt}

\section{Conclusion}
In this work, we reframed in-the-wild AVER as the parameter-efficient and prior-preserving adaptation of language-supervised foundation models. We introduced CLAIP-Emo, a framework that strategically adapts frozen CLIP and CLAP backbones using LoRA (tuning \ensuremath{\le}4\% of parameters), processes temporal information with a tailored asymmetric architecture, and fuses features with a simple fusion head. This efficient design sets a new state of the art on DFEW (80.14\% WAR) and MAFW (61.18\% WAR) with only 8M tunable parameters, surpassing substantially larger, pre-trained models. Our results demonstrate that leveraging language-supervised priors is a scalable and effective alternative to costly domain-specific pre-training for real-world AVER.

\vfill\pagebreak

\small
\bibliographystyle{IEEEbib}
\bibliography{refs}
\end{document}